\documentstyle[prl,aps,multicol]{revtex}
\begin{document}
\title{Reconstructing a pure state of a spin $s$ through three 
         Stern-Gerlach measurements }
\author{Jean-Pierre Amiet and Stefan Weigert }
\address{Institut de Physique de l'Universit\'{e}
    de Neuch\^{a}tel, Rue A.-L. Breguet 1, CH-2000 Neuch\^{a}tel }
\date{August 1998}
\newcommand\ket[1]{|#1\rangle}
\newcommand\bra[1]{\langle #1|}
\newcommand\braket[2]{\langle #1|#2\rangle}
\maketitle
\begin{abstract}

Consider a spin $s$ prepared in a {\em pure} state. It is shown that, generically, the moduli of the $(2s+1)$ spin components along three directions in space determine the state unambigously. These probabilities are accessible experimentally by means of a standard Stern-Gerlach apparatus. To reconstruct a pure state is therefore possible on the basis of $3(2s+1)$ measured intensities. 
\end{abstract}
\begin{multicols}{2}
The reconstruction of a particle density-operator is possible in principle through repeated measurements on an ensemble of identically prepared systems \cite{raymer97,leibfried+98}. Quantum states of vibrating molecules \cite{dunn+95}, of trapped ions \cite{leibfried+96}, as well as the state of atoms in motion \cite{kurtsiefer+97} have been reconstructed successfully in the laboratory. Similarly, quantum optical experiments \cite{smithey+93} have been performed. 

For a spin of length $s$, this question arises for states in a Hilbert space of finite dimension. There is an explicit expression for the density matrix $\rho$ in terms of the moduli of spin components along $(4s+1)$  appropriate directions in space \cite{newton+68}. This number can be reduced to $(2s+1)$ upon adopting a different 
approach \cite{amiet+98}. A standard Stern-Gerlach apparatus with variable orientation in space provides the corresponding probabilities in an experiment. Alternatively, a Wigner function defined on the discrete phase space associated with a finite-dimensional Hilbert space allows one to reconstruct quantum states \cite{leonhardt95}. This method has been adapted in \cite{walser+96} in order to determine a quantized electromagnetic mode
of a cavity. Every proposed method of state reconstruction is bound to reflect on the link between the outcomes of a finite number of measurements obtained in an actual experiment and the mathematical probabilities which refer to {\em infinite} ensembles (see \cite{buzek+96}, for example).    

Suppose now that the spin state to be reconstructed is known to be prepared in a {\em pure} state which is determined by less parameters than a mixed one. How to exploit this additional knowledge in the most efficient way? Reconstruction of pure states has been turned into a question as early as 1933 for a {\em particle} by Pauli \cite{pauli33} who did not provide an answer. One solution of the spin version of the problem \cite{gale+68} makes use of a {\em Feynman filter}. This is an advanced version of a Stern-Gerlach apparatus which is assumed to reveal the relative phases of the expansion coefficients of a pure spin state. Another approach relates expectation values of spin multipoles with the parameters which define the quantum state \cite{bohn91}. 

As shown in this letter, the pure state of a spin $s$ is determined unambigously if the {\em intensities} of the spin components are measured along {\em three} axes. Compared to the $(2s+1)$ axes required for a mixed state \cite{amiet+98}, the experimental effort to perform state reconstruction is thus reduced considerably for large spins. Further, this result is  satisfactory from a mathematical point of view since it generalizes an earlier result: the intensities along two {\em infinitesimally close} axes spanning a plane define a unique pure state when complemented by the expectation value of a spin component 
``out of plane'' \cite{weigert92}. Effectively, this means to measure $(2s+1)$ probabilities along a third direction. 
 
The states of a spin of magnitude $s$ live in a Hilbert space  ${\bf H}^s$ of complex dimension $(2s+1)$, which carries an irreducible representation of the group $SU(2)$. The components of the spin operator $\vec{\bbox{S} }\equiv \hbar \vec{\bbox{s}}$ with standard commutation relations $[\bbox{s}_x,\bbox{s}_y]= i\bbox{s}_z, \ldots$ generate rotations about the corresponding axes. The standard basis of the space ${\bf H}^s$ is given by the eigenvectors of the $z$ component of the spin, denoted by $\ket{s,\mu_z}, -s\leq\mu_z\leq s$. The transformation under the anti-unitary time reversal operator $T$ fixes their phases, $T\ket{s,\mu_z}=(-1)^{s-\mu_z}\ket{s,-\mu_z}$. 
When expanded in the $z$ basis ($\mu_k \equiv \mu_z$), 
\begin{equation}
\ket{\psi}=\sum_{\mu_k=-s}^{s}\psi_{\mu_k} \ket{s,\mu_k} \, , \qquad k = x,y,z \, ,
\label{expandortho}
\end{equation}
a pure state is seen to be determined by $(2s+1)$ complex coefficients $\psi_{\mu_z} \equiv \braket{s,\mu_z}{\psi}$. If normalized, rays $\ket{\psi}$ depend on $4s$ real parameters. Two other bases of the space ${\bf H}^s$ are used in Eq.\ (\ref{expandortho}): the sets $\{ \ket{s,\mu_x} \}$ and $\{ \ket{s,\mu_y} \}$ with $ -s\leq\mu_x,\mu_y\leq s$, made up from the eigenvectors of the spin components $\bbox{ s}_x$ and $\bbox{ s}_y$, respectively. Rotations about appropriate axes by an angle $\pi/2$ map them to the $z$ basis:
\begin{equation}
\ket{s,\mu_z} = e^{ -i \pi \bbox{s}_y/2} \ket{s,\mu_x}
                   = e^{ i\pi\bbox{ s}_x/2} \ket{s,\mu_y} \, .
\label{bases}
\end{equation}

A measurement of the intensities $\{ |\braket{s,\mu_z}{\psi}|^2 \} $ does not fix a single state $\ket{\psi}$ since the phases of the coefficients $\psi_{\mu_z}$ remain undetermined. However: \\
{\em a spin state $\ket{\psi} \in {\bf H}^s$ is determined unambiguously if $3(2s+1)$ probabilities}
\begin{equation}
p(\mu_k)=|\psi_{\mu_k}|^2 \, , \quad k=x,y,z \, ,
\label{data}
\end{equation}
{\em are measured with a Stern-Gerlach apparatus along {\rm three} axes not in a plane. For some exceptional states of  measure zero in Hilbert space ${\bf H}^s$, the probabilities $ p(\mu_k)$ might be compatible with a finite number of states.} 

For simplicity, the proof is carried out for orthogonal axes, the generalization being 
straightforward. Measuring with respect to {\em two} axes provides $2(2s+1)$ intensities 
which are usually compatible with a huge number of isolated states, in agreement with the result of \cite{weigert92}: the parameters fullfil nonlinear relations which may have multiple solutions. Enumerating the ensemble of possible  ``partner'' states is complicated, so a distinctive third measurement is included from the very beginning. 
 
It is useful to rephrase the statement at stake differently. According to (\ref{data}) a state  $\ket{\widetilde \psi}$ gives rise to the {\em same} intensities as does $\ket{\psi}$ if its coefficients 
${\widetilde \psi}_{\mu_k} = \braket{s,\mu_k}{\widetilde \psi}$ differ from $ \psi_{\mu_k}$
by phase factors only. Using (\ref{expandortho}) one writes thus 
\begin{equation}
\sum_{\mu_k =-s}^{s} \psi_{\mu_k} e^{i \chi_k (\mu_k) } \ket{s,\mu_k}   
     = \exp [i \chi_k (\bbox{ s}_k) ] \ket{\psi}  
\, ,  
%
\label{phases}
\end{equation}
with three polynomials $\chi_k (\mu) $ of order $2s$  in $\mu$ at most. From now on, 
the index $k$ is understood to take the values $x,y,$ and $z$ throughout.
The coefficients in (\ref{phases}) thus define three states  $\ket{\psi_k} = W^s_k \ket{\psi}$, where $W^s_k = \exp{ [i\chi_k (\bbox{ s}_k)]}$ is a unitary operator diagonal in the $k$ basis. Consequently, a state $\ket{\widetilde \psi}$ compatible with (\ref{data}) exists if and only if there are nontrivial unitary operators $W^s_k$ such that
\begin{equation}
W^s_x\ket{\psi}    = W^s_y\ket{\psi}
                   = W^s_z \ket{\psi}  
              \equiv \ket{\widetilde \psi} \, .
\label{psitilde}
\end{equation}
It will turn out that this relation is satisfied only if the operators $W^s_k$ are multiples of the identity, implying that $\ket{\widetilde \psi}$ and $\ket{\psi} $ represent the {\em same} ray in Hilbert space.

Before turning to the proof, the intensities $p(\mu_k)$ in  (\ref{data}) are represented in a more compact way. Define three functions $m_k(\alpha)$ of a complex variable $\alpha \in \bbox{ C}$ by   
\begin{equation}
m_k(\alpha)= \bra{\psi}U^s_k(\alpha)\ket{\psi} 
             \equiv  \sum_{\mu_k=-s}^s e^{i\mu_k\alpha}p (\mu_k) \, , 
%
\label{contdata}
\end{equation}
where the operator $U^{s}_k(\alpha)=\exp(i\alpha\bbox{ s}_k)$ rotates a state $\ket{\psi}$ about the $k$ axis 
if $\alpha \in \bbox{ R}$. Eq.\ (\ref{contdata}) is inverted easily using the orthogonality of the functions $\exp[-i\mu_k\alpha]$ on the interval $0\leq \alpha < 2\pi$. 

The proof showing that the data (\ref{data}) are sufficient for state reconstruction is divided into five steps. 
($\imath$) A $2^{2s}$ dimensional ``parent'' space ${\cal H}^s$ is introduced which contains the Hilbert space ${\bf H}^s$ of the spin $s$ as a subspace. ($\imath\imath$) To each state $\ket{\psi} \in {\bf H}^s$ an equivalence class of product states $\{ \ket{\Psi} \in {\cal H}^s \}$ is associated. ($\imath\imath\imath$) A natural definition of {\em generic} states emerges for {\em product } states in ${\cal H}^s$ and, {\em a fortiori}, in ${\bf H}^s$. ($\imath v$) An appropriate set of expectation values of the parent states $\ket{\Psi}$ fixes them uniquely. ($v$) Finally, it is shown that all states $\ket{\widetilde \psi}$ satisfying (\ref{psitilde}) have parents in the {\em same} equivalence class as the original $\ket{\psi}$. Consequently, the (generic) state $\ket{\psi}$ is the only one giving rise to the intensities (\ref{data}).    

($\imath$) The $2^{2s}$ dimensional ``parent'' space ${\cal H}^s$ of ${\bf H}^s$ is obtained from tensoring  $2s$ copies of the Hilbert space $\bbox{ C}^2$ of a spin $1/2$: 
\begin{equation}
{\cal H}^s = \bigotimes_{r=1}^{2s}  \, \bbox{ C}_r^2 \, .
%
%
\label{parentspace}
\end{equation}
A basis of $\bbox{ C}^2$ is given by the eigenstates $\ket{\sigma} \equiv \ket{s=1/2,\mu_3 = \sigma/2}, \sigma = \pm 1$, of the third component of the spin $1/2$: $ \bbox{ \sigma}_3\ket{\sigma}=\sigma\ket{\sigma}$.
This choice induces a basis of ${\cal H}^s$ formed by all product states
\begin{equation}
\ket{\{ \sigma_r\} } =   \bigotimes_{r=1}^{2s}  \, \ket{\sigma_r}\, .
\label{inducedbasis}
\end{equation}
The parent space ${\cal H}^s$ decomposes into a subspace ${\cal H}_{{\text{sym}}}^s$ and its complement, 
\begin{equation}
{\cal H}^s = {\cal H}_{\text{sym}}^s 
                \oplus \left( {\cal H}_{{\text{sym}}}^s \right)^\perp \, ,
\label{decomposition}
\end{equation}
where ${\cal H}_{{\text{sym}}}^s$ is spanned by the $(2s+1)$ states obtained from completely symmetrizing those in (\ref{inducedbasis}):
\begin{eqnarray}
\ket{s,\mu_3} & =         & {\cal S}_{2s} \ket{ \{ \sigma_r \} }  \nonumber \\
                         &\equiv & N_{\mu_3}^s 
                                            \sum_{ \{ \sigma_r \} } \delta(\sigma_1+\cdots + \sigma_{2s} - 2\mu_3) \ket{ \{ \sigma_r \} }  \, ,
\label{basisvect}
\end{eqnarray}
where $-s\leq\mu_3\leq s$, using a symmetrizer of $2s$ objects, ${\cal S}_{2s}$, and the normalization factor 
$N_{\mu_3}^s = ((s-\mu_3)!(s+\mu_3)!/ (2s)! )^{1/2}$. The space ${\cal H}_{{\text{sym}}}^s$ is important here because it carries a $(2s+1)$ dimensional 
irreducible representation of the group of rotations, $SU(2)$, obtained upon reducing the product representation \cite{streater+64}   
\begin{equation}
{\cal U} \, \ket{ \{ \sigma_r \} } =  \bigotimes_{r=1}^{2s} 
                            \sum_{\sigma'_r=\pm1}\ket{\sigma'_r}\bra{\sigma'_r}u_r\ket{\sigma_r} \, ,
\label{repsu2}
\end{equation}
where $u_r$ is the $r$-th copy of a rotation $u \in SU(2)$ of the fundamental representation acting on $\bbox{ C}^2$, and ${\cal U}$ is an operator defined on ${\cal H}^s$. Since Hilbert spaces of the same dimension are isomorphic, ${\cal H}_{\text{sym}}^s$ and ${\bf H}^s$ will be identified from now on.

($\imath \imath$) There is a one-to-one relation between states $\ket{\psi} \in {\cal H}_{\text{sym}}^s$ and equivalence classes of {\em product} states $ \ket{\Psi} \in {\cal H}^s$:
\begin{equation}
\ket{\Psi} \equiv \ket{ \{ \Psi^r \} }
           =  \bigotimes_{r=1}^{2s} 
                 \left(\sum_{  \sigma_r  } \Psi^r_{\sigma_r}\ket{\sigma_r}\right)  \, .
\label{expandparent}
\end{equation}
The equivalence relation $\sim$ is defined as follows: the projection 
of a  state $\ket{ \Psi}$ in (\ref{expandparent}) onto a basis state $\ket{s,\mu_3} \in {\cal H}_{\text{sym}}^s$ must equal the corresponding expansion coefficient of $\ket{\psi}$ 
in the $z$ basis, i.e.\ ,
\begin{equation}
\braket{s,\mu_3}{\Psi} = N_\psi \braket{s,\mu_z}{\psi} \, , \quad -s \leq \mu_3 = \mu_z \leq s \, , 
\label{parentdefined}
\end{equation}
and the factor $N_\psi>0$ may depend on the state $\ket{\psi}$ under consideration but {\em not} on the index $\mu_z$. 
Thus, $\ket{\Psi} \sim \ket{\Psi^\prime}$ means that for a fixed $\ket{\psi}$, the Eqs.\ (\ref{parentdefined}) hold for both product states, $\ket{\Psi} \sim \ket{\Psi^\prime}$. The association of spin states $\ket{\psi}$ with product or ``parent'' states $\ket{\Psi}$ is essential for the following.

In order to determine the class of states satisfying Eq.\ (\ref{parentdefined}) for a prescribed vector $\ket{\psi}$ (with definite phase), multiply by the factor $1/N_{\mu}^s$, by  powers $(-z)^{\mu+s}$ and sum all terms. The right-hand-side then defines an analytic function 
\begin{equation}
f_R(z) = N_{\psi} \sum_{\mu=-s}^s \frac{(-z)^{\mu+s}}{N_{\mu}^s} \, \psi_{\mu} \,  
         \propto \prod_{r=1}^{2s} (z_r - z) \, ,
\label{otherfunctionfz}
\end{equation}
specified by the location of its $2s$ zeroes $z_r$ in the complex plane. The left-hand-side yields a second analytic function of $z$,
\begin{eqnarray}
f_L(z) & =    &    \sum_{\mu=-s}^s\sum_{ \{ \sigma_r\} } (-z)^{\mu+s} 
                     \delta(\sigma_1+\cdots+\sigma_{2s} - 2\mu) 
                               \Psi^1_{\sigma_1}\ldots \Psi^{2s}_{\sigma_{2s}}  \nonumber \\
       &\equiv& \prod_{r=1}^{2s}( \Psi^r_- -  z \, \Psi^r_+ )\, , \qquad  \Psi^r_{\pm} \equiv \Psi^r_{\pm1} \, .
\label{functionfz}
\end{eqnarray}
The $(2s+1)$ equations (\ref{parentdefined}) are satisfied if $f_L(z)$ and $f_R(z)$ coincide. Being two polynomials of degree $2s$, this requires them to have identical zeroes,  
\begin{equation}
\frac{\Psi^r_-}{\Psi^r_+} = z_r \, , \qquad r = 1, \ldots, 2s \, ;
\label{coeffs}
\end{equation}
in addition, $f_L(0)=f_R(0)$ must hold. 
Due to the normalization $\braket{\Psi^r}{\Psi^r} = |\Psi^r_+|^2 + |\Psi^r_-|^2 = 1 $, one can write
\begin{equation}
 \left(\begin{array}{cc}
    \Psi^r _+  \\ 
    \Psi^r_-
       \end{array}
\right) 
 = 
\frac{e^{i\kappa_r}}{\sqrt{1+|z_r|^2}} 
 \left(\begin{array}{cc}
       1\\z_r
       \end{array}
\right) \, ,
\quad \kappa_r \in [ 0, 2\pi ) \, .
%
%
\label{finalcoeff}
\end{equation}
Thus, there are $2s$ undetermined phase factors $e^{i\kappa_r}$ with a product equal to $1$ (remember that  $\ket{\psi}$ denotes a {\em vector}). However, the overall ambiguity is even larger: when comparing the zeroes of the functions $f_L(z)$ and $f_R(z)$, there is no rule 
which would indicate what order to choose when writing down the product state $\ket{\{ \Psi^r \} }$. In other words, the equivalence class of states defined by (\ref{parentdefined}) consists of all states with coefficients (\ref{finalcoeff}) distributed 
in any order over the $2s$ spinors in (\ref{expandparent}). All these states are parents of the same $\ket{\psi}$ since they satisfy Eq.\ (\ref{parentdefined}).     

A given product state $\ket{\Psi}$ with components 
\begin{equation}
\braket{ \{\sigma_r \} } {\Psi} = \Psi_{ \{ \sigma_r \} } = \prod_{r=1}^{2s}\Psi^r_{\sigma_r} \, ,
\label{parentcoe}
\end{equation}
has a unique ``daughter'' $\ket{\psi}$ to be read off directly. Upon parametrizing each factor $\ket{\Psi^r}$ by a complex number $z_r $,
\begin{equation}
 \left(\begin{array}{cc}
    \Psi^r _+  \\ 
    \Psi^r_-
       \end{array}
\right) 
 = 
\frac{1}{\sqrt{1+|z_r|^2}} 
 \left(\begin{array}{cc}
       1\\z_r
       \end{array}
\right) \, ,
\label{parameters}
\end{equation}
one sees that the ensemble $\{ z_r \} \equiv (z_1,...,z_{2s})$ ({\em no} order implied) defines the daughter $\ket{\psi}$ completely while a maximum of $(2s)!$ different parent states $\ket{\Psi}$ is associated with a given set $\{ z_r \}$. 

($\imath\imath\imath$) Suppose that three ensembles of $2s$ real numbers each, $ \{ x_r \} $, $ \{ y_r \} $, and $\{ | z_r | \} \equiv (|z_1|,...,|z_{2s}|)$ with $ z_r = x_r + i y_r$ are given in disorder.
If one is able to construct the disordered ensemble of $2s$ complex numbers $ \{ z_r = x_r + i y_r \}$ upon using the $2s$ conditions $ |z_r|^2 = x_r^2 + y_r^2 $, the equivalence class with representant $\ket{\Psi}$ is called {\em generic}. In other words, it must be possible to combine unambiguously real and imaginary parts into complex numbers $z_r$. In this spirit, a daughter $\ket{\psi} \in {\cal H}_{{\text{sym}}}^s$ will be called {\em generic} if it has generic parents $\ket{\Psi}$. The procedure does not work if equalities such as $x_r = \pm y_{r^\prime}, r\neq r^\prime$ exist; hence {\em exceptional} states have measure zero.  

($\imath v$) It is shown now that the expectation values of rotations ${\cal U}_k (\alpha )$ about the axes $x,y,$ and $z$, fix generic product states $\ket{\Psi}=\ket{\{ \Psi^r \} } $ up to a permutation of the factors $\ket{ \Psi^r} $ and an overall phase factor. A generic $\ket{\Psi} \in {\cal H}^s$ leads to three expectation values 
\begin{equation}
M_k(\alpha) = \bra{\Psi} \, {\cal U}_k(\alpha)\ket{\Psi}
            \equiv \prod_{r=1}^{2s} \bra{\Psi^r} u_k^r (\alpha) \ket{\Psi^r} 
 \, ,
%
\label{genericexp}
\end{equation}
where  $u_k(\alpha)=\bbox{ 1} \cos(\alpha/2) + \bbox{ \sigma}_k \sin(\alpha/2)$ represents 
a rotation about axis $k$ in $\bbox{ C}^2$. Using the parametrization of Eq.\ (\ref{parameters}), the functions $ M_k(\alpha) $ defined in (\ref{genericexp}) read explicitly
\begin{mathletters}
\label{newfunctions:all}
\begin{equation}
M_x(\alpha)  = \prod_{r=1}^{2s}\frac{\cos(\alpha/2)+ 2 i x_r \sin(\alpha/2)}{1+|z_r|^2} \, , 
\label{newfunctions:a}
\end{equation}
\begin{equation}
M_y(\alpha)  =  \prod_{r=1}^{2s}\frac{\cos(\alpha/2)+ 2   y_r \sin(\alpha/2)}{1+|z_r|^2} \, , 
\label{newfunctions:b}
\end{equation}
\begin{equation}
M_z(\alpha)  =  \prod_{r=1}^{2s}\frac{\cos(\alpha/2)+   i (1-|z_r|^2)\sin(\alpha/2)}{1+|z_r|^2}\, ,
\label{newfunctions:c}
\end{equation}
\end{mathletters}
where again $z_r = x_r +i y_r$. Denote by $\ket{{\widetilde \Psi}} \equiv \ket{\{ {\widetilde \Psi}^r \} } $ another product state with expectations ${\widetilde M}_k(\alpha)$:
\begin{equation}
{\widetilde M}_k(\alpha) 
      = \bra{{\widetilde \Psi}}{\cal U}_k(\alpha)\ket{{\widetilde \Psi}} 
      \equiv \prod_{r=1}^{2s} \bra{{\widetilde \Psi}^r} u_k^r (\alpha) 
                      \ket{{\widetilde \Psi}^r}    \, .   
\label{otherPsi}
\end{equation}
Upon describing the state $\ket{{\widetilde \Psi}}$ by the sequence $ \{ {\tilde z}_r \} $, the three functions ${\widetilde M}_k(\alpha)$ are given by Eqs.\ \ref{newfunctions:all}) after replacing each $z_r$ by  ${\tilde z}_r$. It is shown now that the conditions 
\begin{equation}
\bra{{\widetilde \Psi}} \, {\cal U}_k(\alpha)\ket{{\widetilde \Psi}} 
= \bra{\Psi} \, {\cal U}_k(\alpha)\ket{\Psi} 
    \, , \qquad k=x,y,z \, .
\label{equalexp}
\end{equation}
necessitate $\ket{{\widetilde \Psi}} \sim \ket{\Psi} $. Being analytic in the complex $\alpha$ plane, the 
functions $M_k(\alpha)$ and ${\widetilde M}_k(\alpha)$ are equal if they have same zeroes. The equality ${\widetilde M}_z(\alpha)=M_z(\alpha)$ requires $|{\tilde z}_r|=|z_r|$. The condition ${\widetilde M}_x(\alpha)=M_x(\alpha)$ in turn implies ${\tilde x}_r=x_r$; finally, ${\tilde y}_r=y_r$ follows from ${\widetilde M}_y(\alpha)=M_y(\alpha)$. However, this procedure 
determines the ensembles $\{ x_r \}, \{ y_r \}$, and $\{ |z_r| \} $ {\em without} any order of its members. Nevertheless, one can reconstruct the ensemble $\{ z_r \} $ ({\em no} order implied) according to 
($\imath\imath\imath$) if $\ket{\Psi}$ is {\em generic} providing thus a {\em unique} equivalence class. For exceptional states, the $2s$ complex numbers cannot be reconstructed unambigously since they might allow for parents contained in different equivalence classes.      

($v$) The results ($\imath$) to ($\imath v$) imply that the probabilities $p(\mu_k)$ for three directions $x,y$, and $z$ as given in Eq.\ (\ref{data}) determine a generic state $\ket{\psi}$ unambigously. According to Eq.\ (\ref{psitilde}), a state $\ket{{\widetilde \psi}}$ gives rise to the same probabilities as does $\ket{\psi}$ if one has 
\begin{equation}
\ket{\psi_x} = \ket{\psi_y} 
             = \ket{\psi_z}
             = \ket{{\widetilde \psi}}\, .
\label{existWs}
\end{equation}
For parent states $\ket{{\Psi_k}}$ of $\ket{\psi_k}$ this relation says that
\begin{equation}
\ket{{\Psi_x}} \sim \ket{{ \Psi_y}} 
               \sim \ket{{ \Psi_z}}
               \sim \ket{{ \widetilde \Psi}} \, .
\label{existWsparents}
\end{equation}
This implies that the mean values $\bra{\Psi_k} \, {\cal U}_x(\alpha)\ket{\Psi_k}$ of the operator ${\cal U}_x(\alpha)$ = $\otimes_r \exp[i\alpha\bbox{ \sigma}_x/2]$ are equal for $k=x,y,z$: as products they
are invariant under a permutation of their factors. This also holds for expectation values of 
the operators ${\cal U}_y(\alpha)$ and ${\cal U}_z(\alpha)$. Write the parent states $\ket{\Psi_k}$
in the form ${\cal W}_k \ket{\Psi}$ with operators ${\cal W}_k (\{ \alpha_{k,r} \}) = \otimes_r \exp[i\alpha_{k,r} \bbox{ \sigma}_k/2]$ defined on the parent space ${\cal H}^s$ such that they have $W^s_k$ as component acting in ${\cal H}^s_{\text{sym}}$. Contrary to the rotations ${\cal U}_k(\alpha)$ which depend linearly on the generators $\bbox{s}_k$, the operators $W^s_k$ are {\em nonlinear} functions $\chi(\bbox{s}_k)$ of them, Eq.\ (\ref{phases}). Therefore, the operators ${\cal W}_k (\{ \alpha_{k,r} \})$ depend on a set of $2s$ {\em different} angles $\{ \alpha_{k,r} \}$. Using (\ref{existWsparents}) one concludes  
\begin{eqnarray}
\bra{\widetilde \Psi} \,  {\cal U}_k \ket{\widetilde \Psi}  & = &   \bra{\Psi_k}  \,  {\cal U}_k \ket{\Psi_k} \nonumber  \\  
 &  = &  \bra{\Psi} {\cal W}_k^\dagger  \,  {\cal U}_k {\cal W}_k  \ket{\Psi}                    =      \bra{\Psi}  \,  {\cal U}_k  \ket{\Psi}  \,  .
\label{thingscommute}
\end{eqnarray}
The third equality follows because ${\cal W}_k$ and ${\cal U}_k$ do commute, both being functions of $\bbox{s}_k$ only. Eq.\ (\ref{thingscommute}) comes down to saying that the functions $M_k(\alpha)$ and  ${\widetilde M}_k(\alpha)$ coincide for all 
$k$ and $\alpha$.  One concludes thus with ($\imath v$) that the state $\ket{{\widetilde \Psi}}$, a parent of $\ket{{\widetilde \psi}}$, is necessarily a member of the {\em same} equivalence class as the parent $\ket{\Psi}$ of $\ket{\psi}$. In other words, the application of the operators ${\cal W}_k$ on a parent $\ket{\Psi}$ does not map it into another equivalence class. In the generic case, there is thus no state different from $\ket{\psi}$ with the same data (\ref{data}) what was to be shown.

The reasoning ($\imath$) to ($v$) remains valid if one measures the intensities along directions characterized by unit vectors ${\bf n}_\zeta, {\bf n}_\eta$, and ${\bf n}_\xi$  instead of three orthogonal axes. These vectors must be linearly independent, that is, they have to span a {\em volume} in space: ${\bf n}_\zeta \cdot {\bf n}_\eta \times {\bf n}_\xi \neq 0$. 
  
As a matter of fact, it is not excluded that the set of data (\ref{data}) be also sufficient to determine exceptional states unambiguously. Suppose that the numbers 
$\{ z_r \}$ are associated with a parent state $\ket{\Psi}$ and $\{  z_r^\prime \}$ with another one, $\ket{\Psi^\prime}$, where both sets of complex numbers are 
obtained from the ensembles $\{ x_r \}$ and $\{ y_r \}$ through $ |z_r|^2 = x_r^2 + y_r^2 $. This does not necessarily imply the existence of an independent $\ket{\psi^\prime} \neq \ket{\psi}$ since it is the basic conditions 
$| \psi_{\mu_k}^\prime | = | {\psi}_{\mu_k}|$ which must be satisfied. Explicit calculations for low 
values of spin $s$ show that this happens only if ${\psi}_{\mu_k}^\prime = \psi_{\mu_k}^*$, resulting 
in  $\bra{\psi}(\bbox{ s}_y)^{2n+1}\ket{\psi}\equiv 0$ for all integers $n$. In any case one expects every 
non-genericity to vanish if the spatial directions involved are slightly modified.

To sum up, state reconstruction is possible if based on the $3(2s+1)$ moduli of the spin components with respect to 
three directions in space not all in the same plane. Compared to a constructive method using $(2s+1)^2$ real numbers, the non-constructive method presented here requires that considerably less parameters be determined experimentally, namely $3(2s+1)$. 
\end{multicols}
\end{document}